\documentclass[pre,twocolumn,aps]{revtex4}
\usepackage[dvips]{graphicx}
\usepackage{amsmath}
\usepackage{amssymb}
\usepackage[nice]{nicefrac}
\usepackage{color}

\begin{document}

\title{On relation between generalized diffusion equations and subordination schemes}

\author{A. Chechkin}
\email{chechkin@uni-potsdam.de}
\affiliation{Institute of Physics and Astronomy, Potsdam University, Karl-Liebknecht-Strasse 24/25, 14476 Potsdam-Golm, Germany}

\affiliation{Akhiezer Institute for Theoretical Physics, 
Akademicheskaya Str. 1, 61108 Kharkow, Ukraine}

\author{I.M. Sokolov}
\email{igor.sokolov@physik.hu-berlin.de}
\affiliation{Institut f\"ur Physik and IRIS Adlershof, Humboldt Universit\"at zu Berlin, Newtonstra\ss e 15, 12489 Berlin, Germany}

\begin{abstract}
Generalized (non-Markovian) diffusion equations with different 
memory kernels and subordination schemes based on random 
time change in the Brownian diffusion process are popular mathematical 
tools for description of a variety of non-Fickian diffusion processes
in physics, biology and earth sciences. Some of such processes (notably, the fluid limits of continuous time random walks) allow for either
kind of description, but other ones do not. In the present work 
we discuss the conditions under which a generalized diffusion equation 
does correspond to a subordination scheme, and the conditions under which a subordination scheme does possess the corresponding generalized
diffusion equation. Moreover, we discuss examples of random processes 
for which only one, or both kinds of description are applicable. 
\end{abstract}

\maketitle

\section{Introduction}
In his seminal paper of 1961 Robert Zwanzig has introduced a generalized non-Markovian Fokker-Planck equation \cite{Zwanzig}
with a memory kernel, a GFPE in what follows.
The work was much cited, due to the fact that the approach to the derivation of this equation based on the 
projection operator formalism has found its application in the variety of problems in non-equilibrium statistical physics
\cite{Grabert}. The equation itself was however hardly used, except for obtaining Markovian approximations.  
The situation changed when the generalized Fokker-Planck equations with power-law memory kernels gained 
popularity in different fields. This kind of GFPEs, called fractional Fokker-Planck equations (FFPEs) describe the continuous 
(long time-space) limit of continuous time random walks (CTRW) with 
power-law waiting times \cite{Hilfer,Compte,MeerScheff2019}, which gives a physical foundation and explains broad applicability of 
FFPE. Such equations proved useful for description of anomalous transport processes in different 
media. Applications range from charge transport in amorphous semiconductors to underground water pollution and motion of subcellular units in biology, see, e.g., the reviews \cite{MetzlerKlafter,MeKla-2,PhysToday} and references therein, as well as the Chapters in collective monographs \cite{AnoTrans, FracDyn}. Further important generalizations involve kernels consisting of mixtures of power laws, which correspond to distributed-order fractional derivatives \cite{CheGorSok2002, ChechkinFCAA, CheKlaSok2003, APPB,Naber2004,SoKla2005, SokChe2005, UmaGor2005, MeerScheff2005, MeerScheff2006, Langlands,  Hanyga, MaiPag2007, MaiPaGo2007,
CheGorSok2008, Kochubei2008, Meer2011, CheSoKla2012}, or truncated (tempered) power laws \cite{SokCheKlaTruncated, Stanislavsky1, MeerGRL, Baeumer}. Other kernels in use include combinations of power laws with Mittag-Leffler functions and with generalized Mittag-Leffler
functions (Prabhakar derivatives) \cite{TriChe2015,StanWer2016,Trifce1,Trifce2,StanWer2018,
Trifce3,StanWer2019}.

It is well-known that the Markovian, "normal" Fokker-Planck equation, can be obtained from the Langevin equation, the stochastic differential equation for a Brownian motion under the action of an external force \cite{Chandra}. Similarly, the FFPE follows from two Langevin equations, 
giving a parametric representation of the time dependence of the coordinate. These equations describe the evolution 
of the coordinate and of the physical time in an internal time-like variable (operational time) \cite{Fogedby,Baule1,Baule2,Kleinhans,Hofmann,Trifce4}. 
Such an approach is closely related to the concept of subordination, i.e. random time 
change in a random process: A process $X(\tau(t))$ is said to be subordinated to the process $X(\tau)$ under
operational time $\tau(t)$ being a random process with non-negative increments \cite{Feller}. 
The FFPEs discussed above thus describe the Brownian 
motion, possibly in a force field, under a random time change, i.e. a process subordinated to a Brownian motion with or without drift 
\cite{Meerschaert1,Stanislavsky2, Gorenflo1,Gorenflo2,Gorenflo}.
Subordination schemes not only deliver a method of analytical solution of FFPE \cite{SaiZasl,Eli,Meerschaert4,Meerschaert2} or GFPE \cite{SokSub} by its integral transformation to the usual, Markovian, counterpart, but also give the possibility of stochastic simulations of the processes governed by GFPEs \cite{Kleinhans,SaiUt1,MaiPa2003,Piryatinska,MaiPa2006,GoMai2007,Marcin1,Marcin2,Gajda,Meerschaert3,
Stanislavsky3,Annunziato,Marcin3,Marcin4,Stanislavsky4,Stanislavsky5}.

The subordination approach was also used in \cite{BNG, Vittoria1,Vittoria2} in describing, within the diffusive diffusivity model, a recently discovered but widely spread phenomenon of Brownian yet non-Gaussian diffusion, a kind of diffusion process in which the mean squared displacement (MSD) grows linearly
in time, like in normal, Fickian diffusion, but the probability density of the particles' displacements shows a (double-sided)
exponential rather than Gaussian distribution, at least at short or intermediate times \cite{Wang1,Wang2,Chub,Sebastian,Cherail,Grebenkov1,
Grebenkov2,Korean,Sandalo,RalfEPJB}.

This broad use of generalized (not necessarily fractional) Fokker-Planck equations on one hand, and of random 
processes subordinated to the Brownian motion, on the other hand, urges us to put a question on the relation 
of these two kinds of description. In other words, the following questions arise:
(i) given a GFPE (with a certain memory kernel), can one find the corresponding subordinator or show that none exists, and
(ii) given a subordination scheme, can one find the corresponding GFPE (if any), 
or show that none exists. 
These two questions are addressed in our paper.

Our main statements are summarized as follows. Not all valid GFPEs correspond to subordination schemes. Not all subordination schemes possess a corresponding
GFPE. In our paper we give criteria to check, whether a subordination 
 scheme possesses a GFPE, and whether a particular GFPE corresponds to a subordination scheme or not.
 We moreover discuss examples of particular stochastic processes of interest by themselves, having one or another description, or both.

\section{Generalized Fokker-Planck equations and subordination schemes}
Let us first present the objects of our investigation: The GFPEs of a specific form, and the two kinds of
subordination schemes as they are discussed in the literature cited above.

\subsection{Generalized Fokker-Planck equations}

In present paper we discuss equations of the form
\begin{equation}
 \frac{\partial }{\partial t}P(\mathbf{x},t)= \hat{\Phi} \mathcal{L} P(\mathbf{x},t) ,
 \label{NMOp}
\end{equation}
with the linear integrodifferential operator $\hat{\Phi}$ acting on time variable, and ${\cal L}$ is a time-independent linear operator
acting on a function of spatial variable(s) $\mathbf{x}$. We note that Eq.(\ref{NMOp}) is the most popular, but not the most general form of such equations; in Ref. \cite{Zwanzig} a more general form was derived. This includes the possible additional
coordinate or time dependence of $\hat{\Phi}$ and ${\cal L}$, respectively. In the case of time dependence of ${\cal L}$ 
or of position-dependence of $\hat{\Phi}$ the operators may not commute. Such situations were discussed e.g. in Refs. \cite{SokKla} and  \cite{Inhomogeneous}, and are not a topic of present investigation.

In the time domain the corresponding equations are always representable as a GFPE
\begin{equation}
 \frac{\partial }{\partial t}P(\mathbf{x},t)= \int_0^t \Phi(t-t') \mathcal{L} P(\mathbf{x},t') dt'
 \label{eq:Rep1}
\end{equation}
where the memory kernel $\Phi(t-t')$ can be a generalized function (contain delta-functions or derivatives thereof).
Sometimes the corresponding equations come in a form
\begin{equation}
 \frac{\partial }{\partial t}P(\mathbf{x},t)=\frac{\partial}{\partial t}\int_0^t M_R(t-t^{\prime }){\cal L}P(\mathbf{x},t^{\prime })dt^{\prime },  \label{NMFPE1}
\end{equation}
or 
\begin{equation}
\int_0^t M_L(t-t^{\prime }) \frac{\partial }{\partial t'}P(\mathbf{x},t') dt^{\prime }= {\cal L}P(\mathbf{x},t),  \label{NMFPE2}
\end{equation}
however, Eqs. (\ref{NMFPE1}) and (\ref{NMFPE2}) can be reduced to Eq.(\ref{eq:Rep1}). Indeed, in the Laplace domain Eq.(\ref{eq:Rep1}) reads
\[
 u \tilde{P}(\mathbf{x},u) - P(\mathbf{x},0) = \tilde{\Phi}(u)\mathcal{L} \tilde{P}(\mathbf{x},t)
\]
with
\begin{equation}
 \tilde{\Phi}(u) = \left\{
  \begin{array}{ll}
   u \tilde{M}_R(u) & \mbox{for the case Eq.(\ref{NMFPE1})} \\
   1/ \tilde{M}_L(u) & \mbox{for the case Eq.(\ref{NMFPE2}),}
  \end{array}
\right.
\label{eq:OpPhi}
 \end{equation}
where $\tilde{M}_{...}(u) = \int_0^\infty M_{...}(t)e^{-ut} dt$.
We note that for special cases of integral kernels corresponding to fractional or distributed-order derivatives Eqs.(\ref{NMFPE1}) and (\ref{NMFPE2}) were called ``modified'' and ``normal'' forms
of generalized Fokker-Planck equation, respectively \cite{APPB}. Essentially, in the most cases the equation can be expressed in the either form, with the left or the right memory kernel,
but sometimes one of the forms is preferable \cite{APPB,SoKla2005,CheSoKla2012,Trifce4}. Finally, one can also consider schemes with integral kernels on both sides, for which case
\[
 \tilde{\Phi}(u) = u \frac{\tilde{M}_R(u)}{\tilde{M}_L(u)} .
\]

From now on we will use one-dimensional notation for $x$. Generalization to higher spatial dimensions is straightforward.
In all our examples we will concentrate mostly on the case of free diffusion
\begin{equation}
 \mathcal{L} = D \frac{\partial^2}{\partial x^2}.
 \label{LBM}
\end{equation}
The coefficient $D$ has a dimension of the normal diffusion coefficient, $[D]=[\mathrm{L}^2/\mathrm{T}]$, the operator 
$\hat{\Phi}$ is therefore dimensionless, and its integral kernel has a dimension of the inverse time. 
We note however that concentration on free diffusion is not a restriction for the generality of 
our approach since it is only about temporal operators and temporal parts of the subordination procedures.

\subsection{Kernel of GFPE uniquely defines MSD in free diffusion}

Here we show that the form of the memory kernel uniquely defines the mean squared displacement (MSD) in free diffusion, and
vice versa (under mild restrictions). Let us consider the MSD in free diffusion (i.e. without external force and in absence of
external boundaries). We multiply both parts of Eq.(\ref{NMOp}) by $x^2$ and integrate over $x$ to get
\begin{equation}
 \frac{d }{d t}\int_{-\infty}^\infty x^2 P(x,t) dx= D \hat{\Phi} \int_{-\infty}^\infty x^2  \frac{\partial^2}{\partial x^2} P(x,t) dx.
 \label{eq:MSD1}
\end{equation}
Integrating the right hand side of Eq.(\ref{eq:MSD1})  by parts twice and assuming the PDF $P(x,t)$ to vanish at infinity together with its first derivative we get  for the r.h.s.
\[
 \int_{-\infty}^\infty x^2  \frac{\partial^2}{\partial x^2} P(x,t) dx = 2,
\]
so that the evolution of the MSD is governed by 
\begin{equation}
 \frac{d}{dt} \langle x^2 (t) \rangle = 2 D \hat{\Phi} 1,
 \label{eq:ddt}
\end{equation}
with operator $\hat{\Phi} $ acting on a numeric constant. Passing to the Laplace representation we obtain
\[
 u \langle x^2 (u) \rangle = 2 D \tilde{\Phi}(u) \frac{1}{u}
\]
where we assumed to start from an initial condition concentrated at the origin, $\langle x^2 (t = 0) \rangle =0$. This 
uniquely defines $\tilde{\Phi}(u)$ via the MSD:
\begin{equation}
 \tilde{\Phi}(u) = \frac{1}{2D} u^2 \langle x^2 (u) \rangle.
 \label{eq:MSDLap}
\end{equation}

Let us consider our first example. If the MSD in free motion grows linearly in time, $\langle x^2 (t) \rangle = 2 D t$, 
we have $\langle x^2 (u) \rangle = 2 D / u^2$ and therefore $\tilde{\Phi}(u)\equiv 1$
(a unit operator, an integral operator with a $\delta$-functional kernel). 
This means that the \textit{only} GFPE leading to the linear growth
of the MSD is a usual, Fickian, diffusion equation, for which the PDF is Gaussian at all times. 
Therefore, the GFPE is \textit{not} a valid instrument to describe the BnG diffusion, contrary to what is claimed in \cite{Korean}.
On the other hand, the subordination schemes of \cite{Chub,BNG,Grebenkov1} do describe the phenomenon. 

Now we can turn to the main topic of our work: which processes can and which can not be described by the GFPEs. To this end we first discuss a specific integral
representation of the solution to a GFPE and its relation to subordination schemes.

\subsection{Subordination schemes}

Let $X(\tau)$ be a random process parametrized by a ``time-like'' variable $\tau$. Let $\tau$ by itself be a random process, parametrized by the physical time, or \textit{clock time}, $t$. The random process $X(\tau)$ is called the \textit{parent process}, 
the random variable $\tau$ is called the \textit{operational time}, and the random process 
$\tau(t)$ the 
\textit{directing process}, or \textit{subordinator}. The process $t(\tau)$ is called the  \textit{leading process} of the subordination scheme, see \cite{Gorenflo} for the consistent explanation of the terminology used.   The process $X(\tau(t))$ is said to be subordinated to $X(\tau)$ under the operational time $\tau$. The properties of $X(\tau)$ and $\tau(t)$ (or alternatively, $t(\tau)$) fully define the properties of the composite process $X(\tau(t))$.

The fact that the variable $\tau$ is time-like means that the directing process $\tau(t)$ preserves the causality: 
from $t_2 > t_1$ it must follow that $\tau(t_2) \geq \tau(t_1)$: the directing process of a subordination scheme is increasing at least in a weak sense, that is possesses non-negative increments. It is moreover assumed that $\tau(0) = 0$:
the count of operational time starts together with switching the physical clock. 

The composite random function $X(\tau(t))$ can be defined in two ways. The first way corresponds to an explicit definition by defining the (stochastic) equations governing $X(\tau)$ and $\tau(t)$. The second way defines the function parametrically, 
so that the equations for  $X(\tau)$ and $t(\tau)$ are given. 

A process subordinated to a Brownian motion with drift is a process whose parent process is defined
by a stochastic differential equation
\begin{equation}
 \frac{d}{d\tau} x(\tau) = F(x(\tau)) + \sqrt{2D} \xi(\tau)
 \label{SDE}
\end{equation}
with white Gaussian noise $\xi(\tau)$, with $\langle \xi(\tau) \rangle =0$, 
$\langle \xi(\tau_1)\xi(\tau_1) \rangle =\delta(\tau_1-\tau_2)$, whose strength is given by a diffusion coefficient $D$,
and with deterministic drift  $F(x(\tau))$ which will be considered absent in our examples concentrating on free diffusion.

As example of the \textit{explicit}, or \textit{direct}, subordination scheme we name the 
minimal diffusing diffusivity model,
\begin{eqnarray*}
\frac{dx(\tau)}{d\tau} &=& \sqrt{2}\xi(\tau) , \\
\frac{d \tau}{dt} &=& D(t),
\end{eqnarray*}
where the random diffusion coefficient $D(t)$ is a squared Ornstein-Uhlenbeck process \cite{BNG}. 

In the \textit{parametric} subordination scheme the dependence of $t(\tau)$ is given, again
in a form of an SDE, or via an additional transform of its solution. 
The classical Fogedby scheme \cite{Fogedby} corresponds to a stochastic differential equation 
\begin{equation}\label{fogedby}
  \frac{d t}{d\tau} = \lambda(\tau)
\end{equation} 
with $\lambda(t)$ being a one-sided L\'evy noise. The clock time given by the solution of this equation is 
$t(\tau) = \int_0^\tau \lambda(\tau') d \tau'$, , and the PDF $q(t,\tau)$ of the process $t(\tau)$ is given by a one-sided $\alpha$-stable L\'evy law, such that its Laplace transform in $t$ variable reads $\tilde{q}(u,\tau) = \exp(-u^\alpha \tau), 0 < \alpha < 1$.
This scheme corresponds to a diffusive limit of CTRW with a power law waiting time distribution. 

The correlated CTRW model of Ref. \cite{Hofmann} is another example of the parametric scheme 
where $t(\tau)$ is obtained by an additional integration of $\lambda(\tau)$ above:
\begin{equation}\label{hofman}
 \frac{d t}{d\tau} = \int_0^\tau \Psi(\tau - \tau') \lambda(\tau') d \tau'.
\end{equation}
The previous case is restored if the kernel $\Psi(\tau)$ is a $\delta$-function. 

The attractiveness of subordination schemes lays in the fact that if the solution for the PDFs of the parent process at a given operational time,  and of the directing process at a given physical time are known, the PDF $P(x,t)$ of the subordinated process 
can be obtained simply by applying the Bayes formula.
Let $f(x,\tau)$ be the PDF of $x=X(\tau)$ for a given value of the operational time $\tau$, and $p(\tau , t)$ the PDF of the operational time for the given physical time $t$. Than the PDF of $x=X(t) = X(\tau(t))$ is given by 
\begin{equation}
 P(x,t) = \int_0^\infty f(x,\tau) p(\tau,t) d \tau,
 \label{IntFSub}
\end{equation}
which in this context is called the integral formula of subordination \cite{Feller}. The PDF $p(\tau,t)$ of the operational time
at a given clock time is  delivered immediately by explicit schemes,
and can be obtained for parametric subordination schemes by using an additional transformation \cite{Baule1,Gorenflo}, see below. 
Note that the PDF $f(x,\tau)$ for a process subordinated to a Brownian motion always satisfies a usual, Markovian Fokker-Planck
equation
\begin{equation}
\frac{\partial }{\partial \tau}f(x,\tau)={\cal L}f(x,\tau),  \label{Mark}
\end{equation}
with $\mathcal{L}f(x,\tau) = -\frac{\partial}{\partial x} F(x)f(x,\tau)+D\frac{\partial^2}{\partial x^2} f(x,\tau)$
by virtue of Eq.(\ref{SDE}) of the subordination schemes.

\section{The sufficient condition for GFPE to have a subordination scheme}

In Ref.~\cite{SokSub} it was shown that the formal solution of the GFPE (\ref{eq:Rep1}) can be obtained in a form of an integral decomposition 
\begin{equation}
P(x,t)=\int_{0}^{\infty }f(x,\tau )T(\tau ,t)d\tau ,
\label{eq:IntSubord}
\end{equation}
where $f(x,\tau )$ is a solution of a Markovian FPE with the same
Fokker-Planck operator ${\cal L}$, Eq.(\ref{Mark}), and for the same initial and boundary conditions. 
Here the function $T(\tau,t)$ is normalized in its first variable and connected with the memory kernel of the GFPE, as it is discussed in this Section below. The corresponding
form of solution was obtained in \cite{SaiZasl,Eli} for the fractional diffusion and Fokker-Planck equations, and was applied to the 
fractional Kramers equation in \cite{BarSil}; its more general discussion followed in \cite{Sok1}. Equation (\ref{eq:IntSubord}) is akin to the integral formula of subordination, Eq.(\ref{IntFSub}). However, the PDF $P(x,t)$ in Eq. (\ref{eq:IntSubord}) 
may or may not correspond to a PDF of a random process 
subordinated to the Brownian motion with drift (as described by the ordinary FPE), 
since $T(\tau ,t)$ may or may not be a conditional probability density of  $\tau$ at time $t$, e.g. $T(\tau ,t)$ may get negative.
In Ref. \cite{SokSub} the kernels corresponding to subordination schemes with non-negative $T(\tau ,t)$ 
were called "safe", while the kernels not corresponding to any subordination scheme, for which $T(\tau ,t)$ oscillate,  were called ``dangerous''. 
For safe kernels the non-negativity of solutions of GFPE corresponding to non-negative initial conditions is 
guaranteed by virtue of Eq.(\ref{IntFSub}).
The sufficient condition for Eq. (\ref{eq:IntSubord}) to correspond to a subordination scheme will be considered later. 

If one assumes that the solution of GFPE (\ref{eq:Rep1}) can be obtained in the form of integral decomposition  (\ref{eq:IntSubord}) and then insert such form in Eq. (\ref{eq:Rep1}), one gets the Laplace transform of the function $\tilde{T}(\tau ,t)$ in its second variable, 
$\tilde{T}(\tau ,u)=\int_{0}^{\infty }T(\tau ,t)e^{-ut}dt$, as \cite{SokSub}
\begin{equation}
\tilde{T}(\tau ,u)=\frac{1}{\tilde{\Phi}(u)}\exp \left[ -\tau \frac{u}{\tilde{\Phi}(u)}\right] .  \label{Tlap}
\end{equation}
This however, does not answer the questions what are the conditions under which such a solution holds and whether it is unique.
Below we  discuss these issues in some detail by presenting an alternative derivation, i.e. explicitly constructing the solution. 

Let us start from our Eq.(\ref{eq:Rep1}) and integrate its both parts over time getting
\begin{eqnarray*}
&& \int_0^t \frac{\partial}{\partial t''} P(x,t'')  dt''  =  P(x,t) - P(x,0)\\
&& \qquad = \int_0^t dt'' \int_0^{t''} \Phi(t'' - t') {\cal L}P(x,t') dt' .
\end{eqnarray*}
Now we exchange the sequence of integrations in $t'$ and $t''$ on the r.h.s., 
\begin{eqnarray*}
&& \int_0^t dt'' \int_0^{t''} K(t'' -t^{\prime }){\cal L}P(x,t^{\prime })dt^{\prime } \\
&& \qquad  = \int_0^t dt'   {\cal L}P(x,t^{\prime })\int_{t'}^t \Phi(t'' -t')dt'',
\end{eqnarray*}
getting the integral form
\begin{equation}
 P(x,t) - P(x,0) = \int_0^t K(t-t') {\cal L}P(x,t^{\prime })dt^{\prime },
 \label{eq:integrated}
\end{equation}
with the integral kernel $K(t) = \int_0^t \Phi(t'') dt''$ whose Laplace transform is equal to $\tilde{\Phi}(u)/u$. 
Using the condition $\tau(t=0)=0$ we substitute the assumed solution form, Eq.(\ref{eq:IntSubord}), into Eq. (\ref{eq:integrated}):
\begin{eqnarray*}
&& \int_0^\infty f(x,\tau ) T(\tau,t) d \tau - P(x,0) =  \\
&& \qquad \int_0^t dt' K(t-t') \int_0^\infty d \tau \mathcal{L} f(x,\tau ) T(\tau,t').
\end{eqnarray*}
Now we use the assumption that $f(x,\tau )$ is the solution of a Markovian Fokker-Planck equation, and make the substitution $\mathcal{L} f(x,\tau ) = \frac{\partial}{\partial \tau} f(x,\tau )$.
We get:
\begin{eqnarray*}
 && \int_0^\infty f(x,\tau ) T(\tau,t) d \tau - P(x,0) = \\
 && \qquad \int_0^t dt' K(t-t') \int_0^\infty d \tau \left( \frac{\partial}{\partial \tau} f(x,\tau )\right) T(\tau,t').
\end{eqnarray*}
Performing partial integration in the inner integral on the r.h.s., and interchanging the sequence of integrations in $t'$ 
and in $\tau$ in the integral which appears in the r.h.s. we arrive at the final expression
\begin{eqnarray*}
&&  \int_0^\infty f(x,\tau ) T(\tau,t) d \tau - P(x,0) =  \\
&& \qquad \int_0^t K(t - t') \left[f(x,\infty) T(\infty,t') - f(x,0) T(0,t')  \right] dt' \\
&& \qquad - \int_0^\infty d \tau f(x,\tau ) \frac{\partial}{\partial \tau} \int_0^t K(t - t') T(\tau, t') dt' .
\end{eqnarray*}
Now we request that the l.h.s. and the r.h.s. are equal at any time $t$ for all admissible functions $f(x,\tau )$ satisfying the 
Fokker-Planck equation $\frac{\partial}{\partial \tau} f(x,\tau ) = \mathcal{L} f(x,\tau )$ irrespective of the particular form of the
linear operator $\mathcal{L}$ and of the boundary and initial conditions. This gives us three conditions:

\begin{eqnarray*}
&&\int_0^\infty f(x|\tau ) T(\tau,t) d \tau = \\
&& \qquad - \int_0^\infty d \tau f(x|\tau ) \frac{\partial}{\partial \tau} \int_0^t K(t - t') T(\tau, t'), \\
&& -P(x,0) = -f(x,0) \int_0^t K(t - t')  T(0,t'), \\
&& 0 = \int_0^t K(t - t') f(x,\infty) T(\infty,t'),  
\end{eqnarray*}
which can be rewritten as conditions on $T(\tau,t)$ only:
\begin{eqnarray}
&& T(\tau,t) = -  \frac{\partial}{\partial \tau} \int_0^t K(t - t') T(\tau, t') dt', \label{eq:first}\\
&& \int_0^t  K(t - t') T(0, t') dt'= 1, \label{eq:second}\\
&& \int_0^t  K(t - t') T(\infty, t') dt'= 0. \label{eq:Cond3}
\end{eqnarray}
In the Laplace domain Eq.(\ref{eq:first}) turns to a simple linear ODE
\[
 - \frac{\tilde{\Phi}(u)}{u}  \frac{\partial}{\partial \tau} \tilde{T}(\tau, u) = \tilde{T}(\tau, u)
\]
whose general solution is 
\[
 \tilde{T}(\tau, u) = C \cdot \exp\left( - \tau \frac{u}{\tilde{\Phi}(u)} \right)
\]
with the integration constant $C$. This integration constant is set by the second equation, Eq.(\ref{eq:second}), which in the Laplace domain reads
\[
  \frac{\tilde{\Phi}(u)}{u}  \tilde{T}(0, u) = \frac{1}{u},
\] 
so that $C = 1/\Phi(u)$, and therefore 
\[
 \tilde{T}(\tau, u) = \frac{1}{\tilde{\Phi}(u)} \exp\left( - \tau \frac{u}{\tilde{\Phi}(u)} \right),
\]
which is our Eq.(\ref{Tlap}). The function $T(\tau,t)$ is normalized in its first variable, which follows by the direct integration of Eq.(\ref{Tlap}):
\begin{equation}
 \int_0^\infty \tilde{T}(\tau,u) d\tau = \frac{1}{u},
 \label{eq:Norm}
\end{equation}
so that its inverse Laplace transform to the time domain is unity:
\begin{equation}
 \int_0^\infty T(\tau,t) d\tau = 1
\end{equation}
for any $t > 0$.

The third condition, Eq.(\ref{eq:Cond3}), is fulfilled automatically provided $T(\infty, t') = 0$ which implies $\tilde{T}(\infty, u) = 0$.
This is e.g. always the case for non-negative kernels $\Phi(t)$ (as encountered in all our examples)
whose Laplace transform $\tilde{\Phi}(u)$ is positive for all $u$. For non-positive kernels the property has to be checked explicitly.

Let us stress again that the solution in form of Eq.(\ref{Tlap}), which, as we have seen, is 
applicable for a wide range of memory kernels $\Phi$, may or may not correspond to some subordination scheme. 
We note, however, the fact that the kernel does not correspond to a subordination scheme does not devaluate the 
corresponding GFPE by itself, and does not mean that this leads to negative probability densities. 
As an example of a ``dangerous'' kernel let us consider a simple exponential kernel, 
$\Phi(t) = r e^{-rt}$ 
(where the prefactor $r$ of the exponential is added to keep the correct dimension of $\Phi$).
For example, a generalized diffusion equation with an exponential kernel,
\begin{equation}
 \frac{\partial }{\partial t}p(x,t)=\int_0^t r e^{-r(t-t^{\prime })} D \frac{\partial^2}{\partial x^2} p(x,t^{\prime })dt^{\prime },
 \label{expkernel}
\end{equation}
is essentially the Cattaneo equation for the diffusion with finite propagation speed, 
i.e. a kind of a telegrapher's equation, as can be seen by taking a derivative of its both sides w.r.t. $t$:
\begin{equation}
  \frac{\partial^2 }{\partial t^2}p(x,t)  = r \frac{\partial}{\partial t}p(x,t) + r D \frac{\partial^2}{\partial x^2} p(x,t). 
  \label{eq:telegraph}
\end{equation}
The solutions to Eq.(\ref{eq:telegraph}) for non-negative initial conditions are known to be non-negative on the whole real line, 
but changing the operator ${\cal L}$ from a diffusion to a more general one (e.g. to diffusion in presence of the constant force) may lead to 
oscillating solutions \cite{SokSub}.

The reason for this, within our line of argumentation, is that the function $\tilde{T}(\tau,u)$ for equation (\ref{expkernel}),
\begin{equation}
 \tilde{T}(\tau, u) = (1 + u r^{-1}) \exp[-\tau u(1 + r^{-1} u)]
\label{Texp}
\end{equation}
is \textit{not} a Laplace transform of a non-negative function. We remind that the function $\tilde{\phi}(u), 0\leq u \leq \infty$, is a Laplace transform of a non-negative 
function $\phi(t)$ defined on the non-negative half-axis, if and only if $\tilde{\phi}(u)$ is completely monotone, i.e. its derivatives satisfy  
\[
 (-1)^n \phi^{(n)} (u) \geq 0
\]
for $n=0,1,2,...$ \cite{Feller}. On the other hand, it is easy to see that the second derivative of  $\tilde{T}(\tau, u)$ changes its sign. 
Moreover, using the mean value theorem, it is not hard to show that the Laplace transform of any non-negative function integrable to 
unity cannot decay for $u \to \infty$ faster than exponentially (see Appendix \ref{Integral}), which is not true for Eq.(\ref{Texp}).

Summarizing the result of this Section, we see that not all GFPEs can correspond to subordination schemes. The \textit{sufficient condition} to have such scheme is the following:
The kernel  $\Phi(t)$ in the GFPE (\ref{eq:Rep1}) is such that the function $\tilde{T}(\tau,u)$ given by Eq.(\ref{Tlap}) is completely monotone as a function of $u$. 
This always corresponds to subordination scheme, for which $T(\tau,t)$ can be interpreted as a probability density function of the operational 
time  $\tau$ for given physical time $t$, $T(\tau,t) \equiv p(\tau,t)$.

\section{What subordination schemes do have a GFPE?}

Let us first perform some simple manipulations
while assuming that the function $T(\tau,t)$ in the integral decomposition formula (\ref{eq:IntSubord})
does correspond to a subordination scheme, and thus has a meaning of the PDF of operational time $\tau$ for a given physical time $t$, $T(\tau,t) \equiv p(\tau,t)$. Then, in the Laplace domain the function $\tilde{p}(\tau,u)$, Eq.(\ref{Tlap}), can be represented as 
\begin{equation}
 \tilde{p}(\tau ,u) = - \frac{d}{d \tau} u^{-1} \exp \left[ -\tau \frac{u}{\tilde{\Phi}(u)}\right],
 \label{eq:Integral1}
\end{equation}
so that in the time domain we have 
\[
 p(\tau,t) = - \frac{d}{d\tau} \int_0^t q(t',\tau) dt',
\]
where the function $q(t,\tau)$ is given by the inverse Laplace transform of 
\begin{equation}
 \tilde{q}(u,\tau) =  \exp \left[ -\tau \frac{u}{\tilde{\Phi}(u)}\right]. 
 \label{eq:Fcap}
\end{equation}
Thus
\begin{equation}
 \int_{\tau_0}^\infty p(\tau ,t_0) d \tau = \int_0^{t_0} q(t, \tau_0) dt.
 \label{eq:Integral2}
\end{equation}
Now we proceed to show that, as discussed already in \cite{Baule1,Gorenflo}, the function $q(t,\tau)$ has a clear physical meaning: this is namely the PDF of 
clock times corresponding to the given operational time $\tau$. We note here that in spite of the fact that Eq.(\ref{eq:Integral2}) was obtained for a specific form of the PDF $p(\tau,t)$ given by Eq.(\ref{eq:Integral1}), it is more general and gives the relation between the PDFs of a (weakly) increasing process and its inverse.

\begin{figure}[h!]
\centerline{\includegraphics[width=0.9\columnwidth]{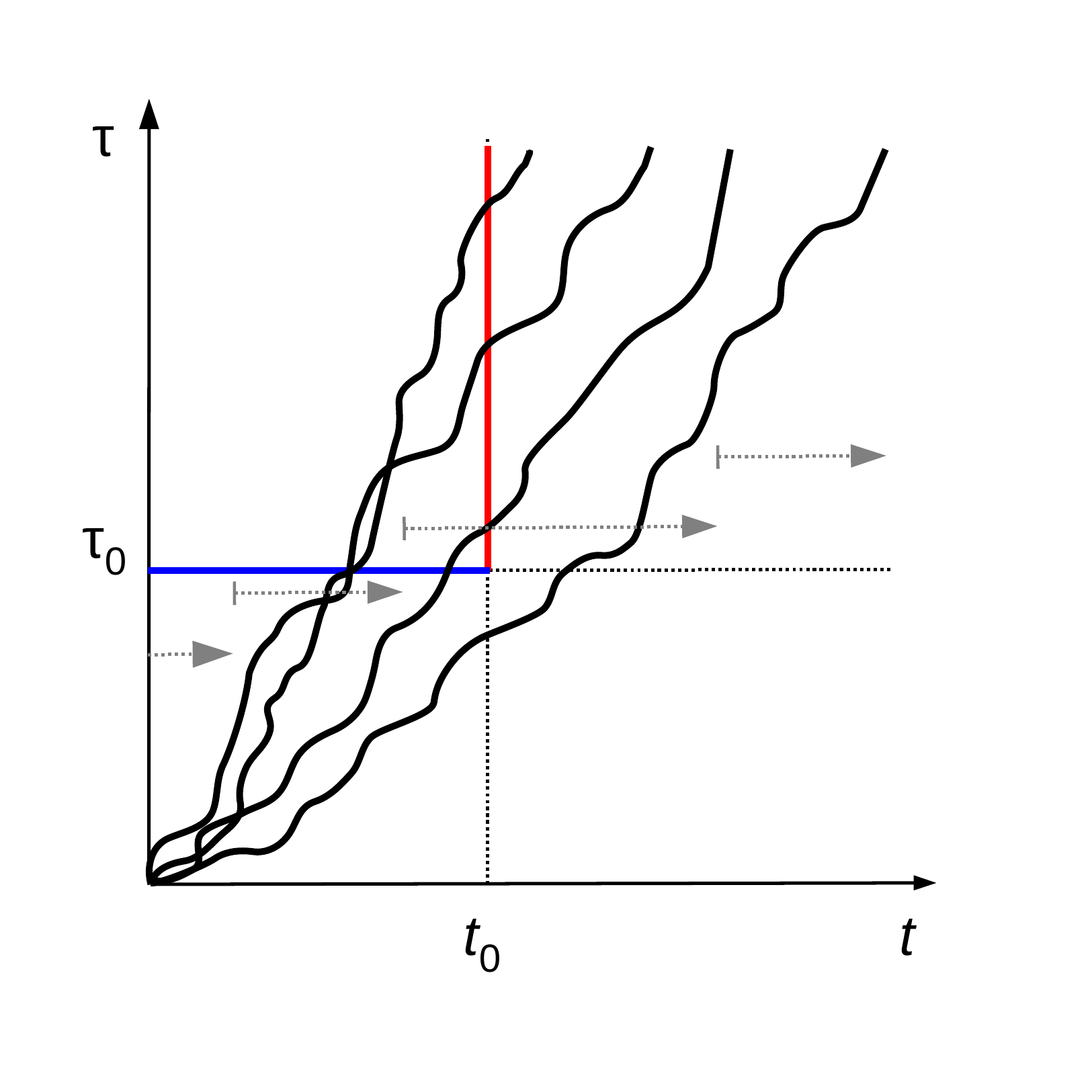}}
\caption{A schematic picture explaining the nature of Eq.(\ref{eq:Integral2}). Black lines: the case of continuous time traces. Here all monotonically non-decaying
traces crossing the horizontal segment $[0, \tau_0)$ (shown in blue online) also cross the vertical one $[\tau_0,\infty)$ (shown in red online), and therefore contribute equally to the r.h.s. and to the l.h.s. of Eq.(\ref{eq:Integral2}).
A gray dashed line shows exemplary a c\`adl\`ag piecewise constant function with jumps, the genuine operational time of a CTRW. 
Any c\`adl\`ag trace passing at a jump from below to above the horizontal segment has to cross the vertical segment during the waiting time. \label{Fig:1}}
\end{figure}

Indeed, let us consider a set of monotonically non-decaying functions, either continuous (diffusive limit) 
or of c\`adl\`ag type (genuine continuous time random walks) on a $(t,\tau)$ plane, see Fig. \ref{Fig:1}. 
The integral on the l.h.s. of Eq.(\ref{eq:Integral2}) counts all functions (with their probability weights) which cross the horizontal segment
$t \in [0, t_0), \tau = \tau_0$, the integral on the r.h.s. counts all functions crossing the semi-infinite vertical segment $t=t_0, \tau \in [\tau_0, \infty)$.
The set of these functions is the same: any monotonically non-decaying function crossing the horizontal segment or passing from its one side to another side on a jump
has to cross the vertical one. No non-decaying function which never crossed the horizontal segment can cross the vertical one. 
Therefore such a monotonicity implies that 
\[
 \textrm{Prob}(\tau > \tau_0 | t_0) = \textrm{Prob}(t < t_0 | \tau_0),
\]
where the probabilities are defined on the set of the corresponding trajectories. The physical meaning of the functions
$\textrm{Prob}(\tau > \tau(t))$ and $\textrm{Prob}(t < t(\tau))$  is that they represent the survival probability 
and the cumulative distribution function for the operational time and for the clock time, respectively. In the continuous case the PDF of a clock time given operational time is then given by 
\begin{eqnarray}
 q(t, \tau) &=& \frac{d}{dt} \textrm{Prob}(t < t(\tau)) = \frac{d}{dt} \textrm{Prob}(\tau > \tau(t)) \nonumber \\ 
&=& \frac{d}{dt} \int_\tau^\infty p(\tau',t) d\tau' .
\label{PDF clock}
\end{eqnarray}
This statement allows for immediate transition between direct (random variable change $\tau(t)$) and parametric (inverse) subordination schemes. This  also gives a necessary condition for a process obeying GFPE to be a subordinated one. 

To obtain such a necessary condition, let us fix two subsequent non-intersecting operational time intervals $\tau_1$ and $\tau_2$ corresponding to 
the physical time intervals $t_1$ and $t_2$. Than $t(\tau_1 + \tau_2) =t(\tau_1) + t(\tau_2) = t_1 + t_2$. Than it follows from Eq.(\ref{eq:Fcap}) that 
\[
\tilde{q}(u, \tau_1 + \tau_2)= \exp \left[ -\tau_1 \frac{u}{\tilde{\Phi}(u)}\right] \cdot 
\exp \left[ -\tau_2 \frac{u}{\tilde{\Phi}(u)}\right],
\]
or, denoting the Laplace characteristic functions of $t_1$ and $t_2$ by $\tilde{\theta}_{t_1}(u)$ and  $\tilde{\theta}_{t_2}(u)$, and the characteristic function of their sum by  $\tilde{\theta}_{t_1+t_2}(u)$ we get
\begin{equation}
 \tilde{\theta}_{t_1 + t_2}(u) = \tilde{\theta}_{t_1}(u) \cdot \tilde{\theta}_{t_2}(u) ,
 \label{eq:subind}
\end{equation}
that is the random variables $t_1$ and $t_2$ are \textit{sub-independent} \cite{Hamedani}.
This property puts a necessary condition for the possibility to describe the subordination scheme by a generalized FPE. The condition is always fulfilled when $t_1$ and $t_2$ are independent (e.g. for a parametric Fogedby scheme \cite{Fogedby}). 

Therefore, the following statement can be made: Only subordination schemes in which the increments of physical time $t(\tau)$ are sub-independent can be described by GFPEs.

Combining Eqs.(\ref{eq:Fcap}) and (\ref{eq:subind}) we find the final criteria for the existence of GFPE for a given subordination scheme.
\begin{itemize}
 \item A direct subordination scheme does posses a GFPE only if the Laplace transform of the PDF $p(\tau,t)$ in its $t$ variable has the form
 \begin{equation}
  \tilde{p}(\tau,u) = \frac{f(u)}{u}\exp(-\tau f(u)).
  \label{criterion1}
 \end{equation}
The kernel of the ensuing GFPE is then $\tilde{\Phi}(u) = u/f(u)$. Here it might be easier to check
that the double Laplace transform has a form
\begin{equation}
 \tilde{\tilde{p}}(s,u) = \int_0^\infty d\tau \int_0^\infty dt e^{-s\tau} e^{-ut} p(\tau,t) = \frac{f(u)}{u[s + f(u)]},
 \label{eq:psu}
\end{equation}
i.e. the function $F(s,u) = 1/ \tilde{\tilde{p}}(s,u)$ is a linear function in $s$: $F(s,u) = a(u) s + 1$.
Using this criterion one can show that the BnG model of Ref. \cite{BNG} does not posses a corresponding GFPE, see Appendix 2. 
\item A parametric subordination scheme does posses a GFPE only if the characteristic function (Laplace transform) of the PDF $q(t,\tau)$ in its $t$ variable has a form
\begin{equation}
 \tilde{q}(u,\tau) = \exp(-\tau f(u)) ,
 \label{criterion2}
 \end{equation}
i.e. the $\tau$-dependence of this function must be simple exponential. The $t$-variables corresponding to non-intersecting $\tau$ intervals are thus sub-independent.  
The kernel of the ensuing GFPE is then $\tilde{\Phi}(u) = u/f(u)$.
\end{itemize}

Using the last criterion one can easily show that there is no GFPE corresponding to correlated CTRW of Ref. \cite{Hofmann}. The distribution of $t$ as a function of $\tau$ 
in this model has a Laplace characteristic function
\begin{equation}
 \tilde{q}(u, \tau) = \exp \left[ - u^\alpha \phi(\tau) \right]
 \label{eq:Hofmann}
\end{equation}
with
\[
 \phi(\tau) = \int_0^\tau d\tau' \left[ \int_{\tau'}^\tau d \tau'' \Psi(\tau'' - \tau') \right]^\alpha
\]
where $\Psi(\tau)$ denotes the memory function for waiting times along the trajectory expressed as a function of the number of steps, cf. Eq.(26) of
 Ref. \cite{Hofmann}. To correspond to any GFPE the argument of the exponential in Eq.(\ref{eq:Hofmann}) has to be linear in $\tau$, i.e. $\phi(\tau) = a \tau$, where $a$=const. This means that the square bracket in the expression for $\phi$
must be a constant (equal to $a$) and therefore $\Psi(\tau) = a^{1/\alpha} \delta(\tau)$, which corresponds to a standard, non-correlated CTRW.

\section{Conclusions}
\
Growing awareness of the complexity of non-Markovian diffusion processes in physics, earth sciences and biology gave rise to a spark of interest to mathematical tools capable to 
describe such non-standard diffusion processes beyond the Fick's law.
Generalized diffusion equations on one hand, and subordination schemes, on the other hand, are the two classes of such instruments, which were successfully used for
investigation of a broad variety of anomalous diffusion processes. For several situations, notably, for decoupled continuous time random walks, both are applicable,
and stand in the same relation to each other as the Fokker-Planck equation and the Langevin equation do for the case of normal, Fickian diffusion. 
In the present work we address the question, whether this is always the case. The answer to this question is negative: some processes described by the generalized
diffusion equations do not possess an underlying subordination scheme, i.e. cannot be described by a random time change in the normal diffusion process. 
On the other hand, many subordination schemes do not possess the corresponding generalized diffusion equation. The example for the first situation is the Cattaneo equation,
which can be represented as a generalized diffusion with exponential memory kernel, for which no subordination scheme exists. The example of the second situation is
the minimal model of Brownian yet non-Gaussian diffusion and correlated CTRW, the subordination schemes for which we show that no corresponding generalized diffusion equation can be put down.
We discuss the conditions under which one or the other description is applicable, i.e. what are the properties of the memory kernel of the diffusion equation sufficient 
for its relation with subordination, and what are the properties of the random time change in the subordination scheme necessary for existence of the corresponding generalized diffusion equation.

\section{Acknowledgements}
AC acknowledges support from the NAWA 
Project PPN/ULM/2020/1/00236.

\appendix

\section{Laplace transform of a non-negative function \label{Integral}}

Let us show that the Laplace transform $\widetilde{\phi}(u)$ of any \textit{non-negative} function $\phi(t)$ integrable
to a constant, i.e. with $\phi(t) \geq 0$ and with $0 < \int_0^\infty \phi(t)dt < \infty$, cannot decay faster than exponentially 
with the Laplace variable $u$:
\[
\widetilde{\phi}(u) = \int_0^\infty \phi(t) e^{-ut} dt \geq A e^{-Bu},
\]
with a positive constant $A$ and a non-negative constant $B$.

To see this we use the following chain of relations:
\begin{eqnarray*}
 \int_0^\infty \phi(t) e^{-ut} dt &\geq& \int_0^C \phi(t) e^{-ut} dt \\
 &=& e^{-Bu} \int_0^C \phi(t) dt = A e^{-Bu}.
\end{eqnarray*}
Here $C$ is some cut-off value which is chosen such that $\int_0^C \phi(t) dt > 0$, which is always possible
due to our assumptions about the integrability of $\phi(t)$ to a positive constant.
The inequality follows from the mean value theorem for integrals: 
by assumptions $\phi(t)$ is non-negative and integralble, and $e^{-ut}$ is,
evidently, continuous. The value of $B$ then follows the inequality $0 \leq B \leq C$, i.e. is non-negative.

For our function $T(\tau,t)$, Eq.(\ref{Texp}), the Laplace transform $T(\tau,u)$ is defined for $u=0$ so that 
$\int_0^\infty T(\tau,t) dt = 1 > 0$ for any $\tau$. Eq.(\ref{Texp}) essentially corresponds to a Laplace transform of a function
strongly oscillating at small $t$.

\section{The minimal model of BNG diffusion}

In the BnG model of Ref. \cite{BNG} the PDF $p(\tau,t)$ (denoted there as $T(\tau,t))$ is defined via its Laplace transform in $\tau$ variable,
\begin{eqnarray*}
\lefteqn{ \tilde{p}(s,t)  =} \\
&& \hspace{-0.5cm} \frac{e^{t/2}}{\sqrt{\frac{1}{2} \left(\sqrt{1+2s} + \frac{1}{\sqrt{1+2s}} \right) \mathrm{sinh}(t \sqrt{1+2s}) + \mathrm{cosh}(t \sqrt{1+2s}) }}. 
\end{eqnarray*}

Let us take the double Laplace transform
\begin{equation}
\tilde{\tilde{p}}(s,u)  = \int_0^\infty \tilde{p}(s,t)  e^{-ut} dt
 \label{eq:Lus}
\end{equation}
and check whether it has the form
of Eq.(32). 

We first denote $\alpha = \sqrt{1 + 2s} > 1$ and rewrite the function $\tilde{p}(s,t) $ as
\begin{eqnarray*}
\tilde{p}(s,t)  &=&  \frac{e^{t/2}}{\sqrt{\frac{1}{2} \left(\alpha + \frac{1}{\alpha} \right) \mathrm{sinh}(\alpha t) + \mathrm{cosh}(\alpha t)}} \\
&=& \frac{2 \sqrt{\alpha} e^{-\frac{\alpha-1}{2}t}}{(\alpha+1)^2} \left[1 + \left( \frac{\alpha-1}{\alpha+1}\right)^2 e^{-2 \alpha t} \right]^{-\frac{1}{2}}.
\end{eqnarray*}
Denoting $A = 2 \sqrt{\alpha}/(\alpha+1)^2$ and $\zeta = \left[ (\alpha-1)/(\alpha+1)\right]^2$ we get
\[
\tilde{p}(s,t)  = A e^{-\frac{\alpha-1}{2}t}\left(1 + \zeta  e^{-2 \alpha t} \right)^{-\frac{1}{2}}.
\]
Substituting this expression into Eq. (\ref{eq:Lus}) and denoting $\beta = (\alpha-1+2u)/2$,  
we obtain
\[
\tilde{\tilde{p}}(s,u)  = A \int_0^\infty \left(1 + \zeta e^{-2 \alpha t} \right)^{-\frac{1}{2}} e^{-\beta t} dt.
\]
Now we change the variable of integration to $x = e^{-2 \alpha t}$ to arrive to the expression
\[
\tilde{\tilde{p}}(s,u)  = \frac{A}{2 \alpha} \int_0^1 \left(1 + \zeta x \right)^{-\frac{1}{2}} x^{\frac{\beta}{2 \alpha}-1} dx.
\]
This can be compared with the integral representation of the hypergeometric function:
\begin{eqnarray*}
 && \;_2F_1(a,b;c;z) = \\
 && \qquad \frac{\Gamma(c)}{\Gamma(b) \Gamma(c-b)} \int_0^1 x^{b-1}(1-x)^{c-b-1}(1-xz)^{-a} dx,
\end{eqnarray*}
Eq.(15.3.1) of Ref. \cite{AbraSteg}, from which we get $a = \frac{1}{2}$, $b = \frac{\beta}{2 \alpha}$, $c = b + 1$, and $z = - \zeta$ so that,
\[
\tilde{\tilde{p}}(s,u)   = \frac{A}{\beta} \;_2F_1 \left(\frac{1}{2},\frac{\beta}{2a},1+ \frac{\beta}{2a}, - \zeta \right).       
\]
Substituting the values of parameters we obtain:
\begin{eqnarray*}
&& \tilde{\tilde{p}}(s,u)  = \frac{4(1+2s)^{\frac{1}{4}}}{(1+\sqrt{1+2s})^2(\sqrt{1+2s}+2u-1)} \times \\
&& \qquad_2F_1 \left[\frac{1}{2}, \frac{\sqrt{1+2s}-1+2u}{4\sqrt{1+2s}}, \frac{5\sqrt{1+2s}-1+2u}{4\sqrt{1+2s}},\right. \\
&& \qquad \qquad  \left. - \left(\frac{\sqrt{1+2s}-1}{\sqrt{1+2s}+1} \right)^2 \right].
\end{eqnarray*}

\begin{figure}[h!]
 \centerline{\includegraphics[width=0.9\columnwidth]{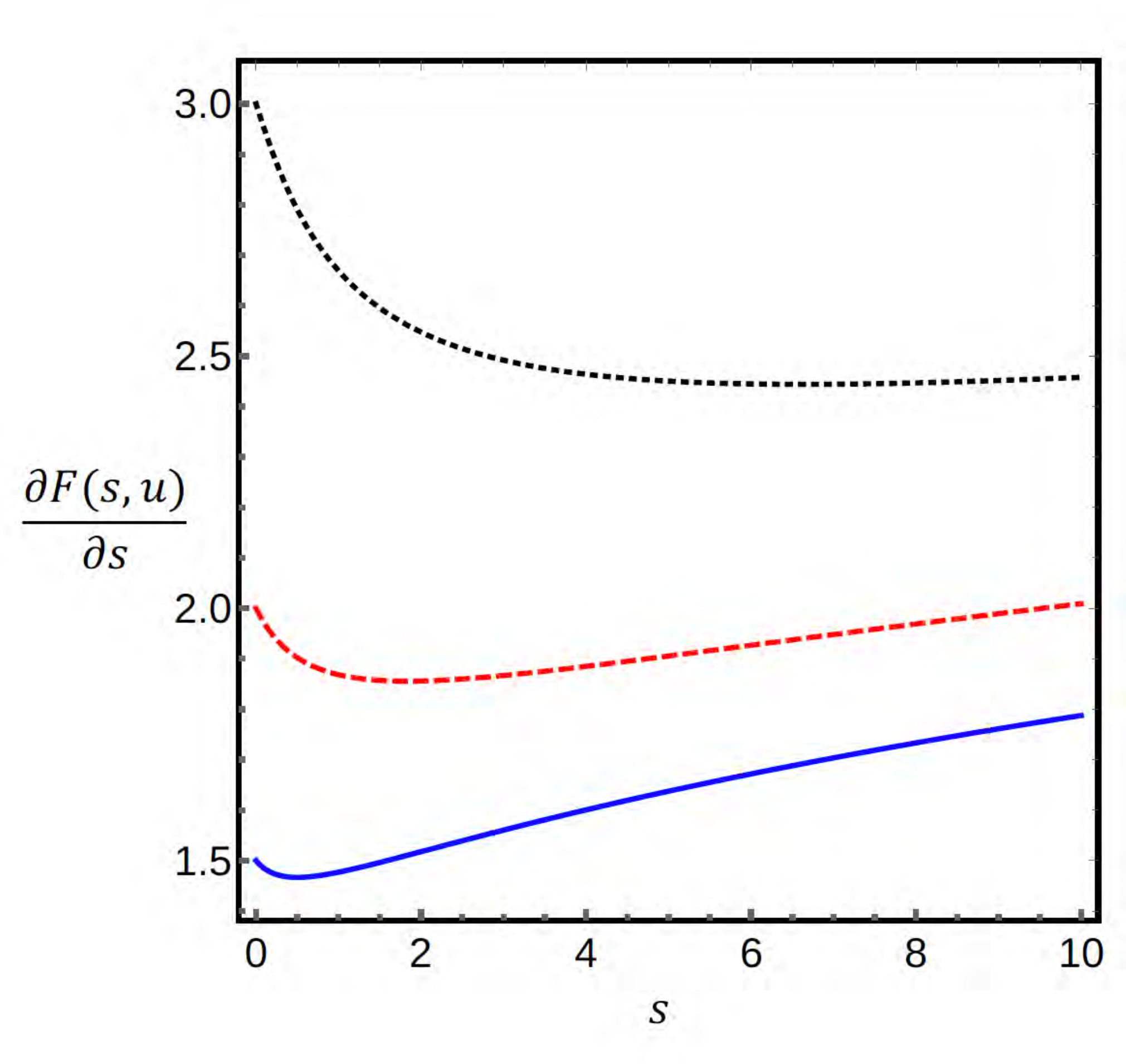}}
 \caption{The $s$-derivative of the function $F(s,u)$ for $u=0.5,1$ and $2$, shown by solid, dashed and dotted lines, respectively. \label{fig:deriv}}
\end{figure}

The function $F(s,u)=1/\tilde{\tilde{p}}(s,u) $ is not a linear function of $s$ for fixed $u$. This can be clearly seen when plotting the $s$-derivative of this function for fixed $u$ with the help of Mathematica, see Fig. \ref{fig:deriv}.

\end{document}